\documentclass[a4]{iopart}
\usepackage{epsfig,graphicx,rotating,pst-all,url}
\usepackage{amssymb}
\usepackage{epic}

\newcommand{\be}{\begin{equation}} 
\newcommand{\ee}{\end{equation}}
\newcommand{\bea}{\begin{eqnarray}}
\newcommand{\eea}{\end{eqnarray}}
\newcommand{\bc}{\begin{center}}
\newcommand{\ec}{\end{center}}

\begin{document}

\title{Angular momentum non conserving symmetries in bosonic models}
\author{L.Fortunato$^1$, W.A. de Graaf$^{2}$}
\address{$^1$ ECT*, Strada delle Tabarelle 286, I-38123 Villazzano (Trento), Italy  \\
$^2$ Dip. Matematica, Universit\`a di Trento, via Sommarive 24, I-38123  Povo (Trento), Italy }

\begin{abstract}
The Levi-Malcev decomposition is applied to bosonic models of quantum mechanics based on unitary Lie algebras $u(2)$, $u(2) \oplus u(2)$, $u(3)$ and $u(4)$ to clearly disentangle semisimple subalgebras. The theory of weighted Dynkin diagrams is then applied to identify conjugacy classes of relevant $A_1$ subalgebras allowing to introduce a complete classification of new angular momentum non conserving (AMNC) dynamical symmetries. The tensor analysis of the whole algebra based on the new "angular momentum" operators reveals unexpected spinors to occur in purely bosonic models. The new chains of subalgebra can be invoked to set up ANMC bases for diagonalization.
\end{abstract}
\pacs{-}

\section{Introduction} 
In this paper we reanalyze some bosonic models, commonly used in various branches of physics, under a new perspective that lead us to explore angular momentum non conserving dynamical symmetries.
As a first step we use the well-known Levi-Malcev decomposition, that allows to establish a clear relation between semisimple and non-semisimple Lie algebras, consistently for unitary algebras arising in bosonic models of quantum mechanics. Although this is often considered a very elementary step, its role in the classification scheme of subalgebras is absolutely fundamental. Considering the crucial importance of these bosonic models \cite{IaAr,IL,Iac,FvI,Oss} in nuclear, molecular and other branches of physics, it seems desirable to set order in this matter by discussing the lowest rank examples first. In addition, when the rank grows, we apply the theory of weighted Dynkin diagrams \cite{CMG} to classify the different conjugacy classes of $A_1$-type subalgebra (Cartan's notation for the isomorphic algebras $sl_2\sim su(2) \sim so(3)$) in terms of weighted Dynkin diagrams. This yields a complete classification of subalgebras that brings in the possibility of additional chains that do not conserve the (usual) angular momentum. A complete tensor analysis of the whole $u(4)$ algebra is carried out with respect to each particular $A_1$ subalgebra: this reveals several interesting aspects, as, for example, the occurrence of spinor operators or the appearance of octupole operators. 
With several additional chains of subalgebras available, one might look for dynamical symmetries and basis states (labeled by quantum numbers of the Casimir operators of the subalgebras in the chain) that do not conserve the usual angular momentum, but rather conserve some quantity that has the formal properties of an angular momentum. This is certainly not a basis that one might have guessed from physical principles, nevertheless the mathematical structure ensures that it exists and it might very well turn out that,  by working with this basis, it could be easier to solve the diagonalization problem.
All the algebras we deal with have also been implemented in the \textsf{GAP4} programming language \cite{GAP4}, that allows symbolic manipulations of algebraic objects with the help of the \textsf{SLA} package \cite{sla} that has been especially designed to provide relevant computational algorithms. In order to ease the control and reproduction of some of the results contained in this paper we provide also simple script files \cite{files} that can be run under \textsf{GAP4}.

The paper is organized as follows:  in section \ref{math} we will introduce the mathematical preliminaries that are necessary for the comprehension of the rest, in particular in subsection \ref{nil} the classification of subalgebras of $sl(n)$ is explained and in subsection \ref{tens} the notion of spherical tensors is briefly reminded; in sections \ref{sec-u2} and \ref{sec-u2u2} we discuss the illustrative cases of $u(2)$ and $u(2)\oplus u(2)$, using a convenient boson representation; in section \ref{sec-u3} we apply the theory of weighted Dynkin diagrams to the case of $u(3)$ and we discuss its tensor analysis; in section \ref{sec-u4} we study the important case of $u(4)$,  that is used in the vibron model of diatomic molecules. We give a complete tensor classification of all the operators in this algebra with respect to all possible classes of $A_1$ subalgebras.  We discuss in section \ref{appl} some physical applications of these findings. In particular we show how the basis states coming from an angular momentum non conserving chain can be used to diagonalize model hamiltonians. Finally, in section \ref{conc} we draw some conclusions and perspectives.

\section{Mathematical preliminaries}
\label{math}
\subsection{Semisimple subalgebras of $sl(n)$}
\label{nil}
The classification of the semisimple subalgebras of a semisimple Lie algebra
is a well-studied topic, see, for example, \cite{dyn,logru,graaf_sss}.
Here we focus on the case of the Lie algebra $sl(n)$, as this is the main case
of interest for the rest of the paper.

Recall that $sl(n)$ is the Lie algebra of $n\times n$-matrices over $\mathbb{C}$ with trace $0$.
The group corresponding to this algebra is $SL(n)$, consisting of the $n\times n$-matrices with
determinant 1. The group acts on the Lie algebra: let $g\in SL(n)$ and $x\in sl(n)$, then
the result of letting $g$ act on $x$ is $gxg^{-1}$. This action sends subalgebras into
isomorphic subalgebras. The classification problem is to get a list of semisimple subalgebras
of $sl(n)$, such that any other semisimple subalgebra can be obtained from an element
of the list, by acting with a suitable element of $SL(n)$. The diagram of Figure \ref{u4} 
displays this classification for $sl(4)$.

Of particular interest in this paper are the subalgebras that are isomorphic to $sl(2)$.
Such a subalgebra has a basis $x,y,h$ with $[h,x]=2x$, $[h,y]=-2y$, $[x,y]=h$. Now after
acting with an element of $SL(n)$ we may assume that $h$ lies in $H$, the subalgebra of
$sl(n)$ consisting of diagonal matrices with trace 0. In other words, $h=\sum_{i=1}^n a_i
e_{i,i}$, where $e_{i,i}$ is the $n\times n$-matrix with a 1 on position $(i,i)$ and zeros
elsewhere, and $\sum_i a_i = 0$. Because of the last consition, $h$ is already determined
by the differences $a_i-a_{i+1}$, for $1\leq i\leq n-1$. It can be shown that, after possibly
acting with another element of $SL(n)$, we get $a_i-a_{i+1}\in \{0,1,2\}$. 

The Dynkin diagram of $SL(n)$ is 

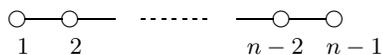
\begin{center}
\begin{picture}(150,20)
  \put(3,10){\circle{6}}
\put(3,-3){{\small 1}}
  \put(23,10){\circle{6}}
\put(23,-3){{\small 2}}
  \put(103,10){\circle{6}}
\put(90,-3){{\small $n-2$}}
  \put(123,10){\circle{6}}
\put(120,-3){{\small $n-1$}}
  \put(6,10){\line(1,0){14}}
  \put(26,10){\line(1,0){14}}
\dashline[20]{2}(50,10)(75,10)
  \put(86,10){\line(1,0){14}}
  \put(106,10){\line(1,0){14}}
\end{picture} 
\end{center}

\noindent To node $i$ we add the label $a_i-a_{i+1}\in \{0,1,2\}$. The result is what is called the
{\em weighted Dynkin diagram}. It can be shown that, up to the action of $SL(n)$, a subalgebra
isomorphic to $sl(2)$ is completely determined by its weighted Dynkin diagram. In other words,
two such subalgebras are conjugate under $SL(n)$ if and only if they have the same weighted
Dynkin diagram. Therefore in this paper we identify a subalgebra isomorphic to $sl(2)$
by its weighted Dynkin diagram (abbreviated: WDD). Furthermore, since we will always be
dealing with the Lie algebras $sl(n)$, we give a weighted Dynkin diagram by the
sequence of its labels: $[a_1-a_2,\ldots,a_{n-1}-a_n]$.

\subsection{Spherical tensors}
\label{tens}
We recall here a few basic concepts and definitions about tensors and tensor products, that are important for the rest of the paper.
The reader who is unfamiliar with this topic might find plenty of material in Ref. \cite{IaAr,IL,Iac,FvI}, that are very close to the subject of the present paper. Several other textbooks of quantum mechanics and group theory introduce this very same subject to various levels of mathematical complexity.

A tensor operator $T^k_q$ is an operator that satisfies the following commutation relations:
\be
[X_i, T^k_q]= \sum_{q'} \langle kq' \mid X_i \mid kq\rangle T^k_{q'}
\ee
with all elements of a certain Lie algebra $g$, i.e. $\forall X_i \in g$. The basis $\mid kq\rangle $ is generically labeled in such a way that $k$ and $q$ identify the irreducible representations of $g$ and of some relevant subalgebra $g'$, respectively. Clearly, for high rank algebras the number and kind of subalgebras is very large and there might be reductions that are not multiplicity free.
Fortunately here we will deal only with the case of tensors with respect to $so(3)\supset so(2)$ (or almost equivalently $u(2) \supset u(1)$), therefore $k$, integer or semi-integer, will be the label of representations of $so(3)$ (i.e. connected to the eigenvalue of the quadratic Casimir operator of $so(3)$), and $q$ will be the label of $so(2)$ with values $q=-k,\dots,k$ in integer steps.
Tensor operators with respect to $so(3)$, also called spherical tensors, satisfy the following commutation relations:
\bea
~[J_z, T^{k}_q] &=& q\; T^{k}_q \\
~[J_\pm, T^{k}_q] &=& \sqrt{(k\pm q+1)(k \mp q)} \;T^{k}_{q\pm 1} \;. \nonumber 
\label{sto}
\eea
where $J_z, J_\pm$ are the operators that form $so(3)$.  This definitions are due to Racah \cite{Rac}.
The rank of the tensor is $k$ and it must not be confused with the rank of the algebra. It is customary to talk about scalar ($k=0$), vector ($k=1$), quadrupole ($k=2$), octupole ($k=3$) operators, etc. One point of the tensor analysis, that is discussed in the next sections, is precisely to attribute a certain tensorial character (i.e. a rank) to all operators that form an algebra {\em with respect to a  certain given subalgebra}.

One can couple tensors to form a new tensor, much in the same fashion in which representations are coupled. Formally, we can define the tensor coupling or product (with respect to $so(3)$ in the present case) as
\be
[T^{k_1} \times U^{k_2}]^k_q= \sum_{q_1,q_2} \langle k_1 q_1 k_2 q_2\mid kq\rangle T^{k_1}_{q_1} U^{k_2}_{q_2} 
\ee
where the two tensors $T$ and $U$ with rank $k_1$ and $k_2$ are coupled to a new tensor of rank $k$, according to the triangular condition $k=\mid k_1-k_2\mid, ..., k_1+k_2$, and the third components $q_1$ and $q_2$ are summed to $q$, according to the addition law for the third components. The coefficients appearing in the sum are simply Clebsch-Gordan coefficients. For the sake of clarity, scalars and scalars can only couple to a scalar ($\vec 0+\vec 0=\vec 0$), vector and scalar couple to a vector ($\vec 0+\vec 1=\vec 1$), vector and vector couple to either a scalar, a vector or a quadrupole tensor ($\vec 1+\vec 1= \{\vec 0,\vec 1,\vec 2 \}$) and so on. Tensors with semi-integer rank are called spinors.

\section{$u(2)$}
\label{sec-u2}
The two dimensional harmonic oscillator problem, that arises frequently in simple models of quantum mechanics \cite{FvI}, is connected to the $u(2)$ Lie algebra by means of two species of (scalar) bosons, called $s$ and $t$ that obey standard bosonic commutation rules of the form $[s,s^\dag]=[t,t^\dag]=1$ and $[s,t]=[s,t^\dag]=[s^\dag,t]=[s^\dag,t^\dag]=0$.
The Lie algebra $u(2)$ is built with bilinear operators made up of these operators and it amounts to four elements:
\be
\mathfrak{g}: \quad g_1=s^\dag s, \quad g_2=s^\dag t, \quad g_3=t^\dag s, \quad g_4=t^\dag t \;.
\label{op}
\ee
Here we had no reason to use the formalism of tensor products because both the building blocks and all the bilinear operators are scalars and the tensor couplings are trivial.
This algebra, $\mathfrak{g}$, is not semisimple. Using the Levi-Malcev decomposition it may be written as the direct sum of a 1-dimensional subalgebra, $\mathfrak{r}$, the radical or maximal solvable ideal, and a 3-dimensional semisimple Levi subalgebra $\mathfrak{s}$, as follows 
$u(2)\simeq ~~\mathfrak{r}\equiv u(1) ~~\oplus ~~\mathfrak{s}\equiv su(2)$, 
with $\mathfrak{r}=\langle g_1+g_4\rangle $ and $\mathfrak{s}=\langle g_2,g_3, g_1-g_4\rangle $. Notice that $\mathfrak{s}$ has a basis consisting of elements of the form given in Eq. (1.20) of Ref. \cite{FvI}, as the standard generators of $su(2)$ angular momentum algebra, namely $\hat J_x=(s^\dag s-t^\dag t)/2,\hat J_y=(t^\dag s+s^\dag t)/2,\hat J_z=i(t^\dag s-s^\dag t)/2$.
Clearly the basis element of the radical is the total number of bosons operator, 
$\hat N= s^\dag s+ t^\dag t$, that is known to commute with all the operators, thereby forming the center of the Lie algebra, i.e. $\mathfrak{r} \equiv \mathfrak{c}$ \footnote{The semisimple part is unique up to inner automorphisms of the form $exp(ad_z)$, where $z \in \mathfrak{r}$. Hence here it happens that $exp([\hat N,x])=1, \forall x \in \mathfrak{s}$ and $\mathfrak{s}$ is unique.}.
Although in the present case it is almost trivial, the semisimple part is amenable to treatment with the theory of weighted Dynkin diagrams \cite{CMG}. It correspond to $A_1$ in Cartan notation and admits only one conjugacy class of semisimple subalgebras, labeled by the weighted Dynkin diagram: $[2]$. 

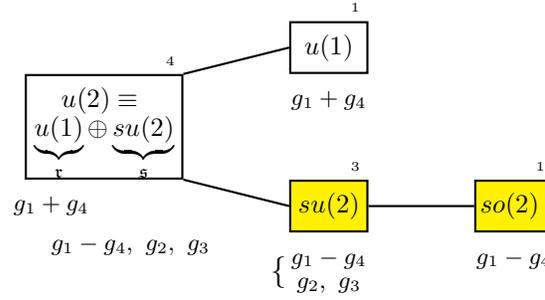
\begin{figure}[!t]
\bc
\begin{picture}(200,100)(0,0)
\psset{unit=1pt}
\psframe(0,80)(60,40)  \rput(55,85){\tiny 4}
\rput(30,70){$u(2)\equiv$ } 
\rput(30,52){ $\underbrace{u(1)}_{\mathfrak{r}}\oplus \underbrace{su(2)}_{\mathfrak{s}}$ }
\psline(60,80)(100,90)
\psline(60,40)(100,30)

\psframe(100,100)(130,80)  \rput(125,105){\tiny 1}
\rput(115,90){ $u(1)$ }
\psframe[fillcolor=yellow, fillstyle=solid](100,40)(130,20) \rput(125,45){\tiny 3}
\rput(115,30){ $su(2)$ }
\psline(130,30)(170,30)

\psframe[fillcolor=yellow, fillstyle=solid](170,40)(200,20)  \rput(195,45){\tiny 1}
\rput(185,30){$so(2)$ }
\rput(10,30){\small $g_1+g_4$}
\rput(40,15){\small $g_1-g_4,~g_2,~g_3$}
\rput(115,70){\small $g_1+g_4$}
\rput(96,5){\large $\{$}
\rput(115,10){\small $g_1-g_4$}
\rput(115,0){\small $g_2,~g_3$}
\rput(185,10){\small $g_1-g_4$}
\end{picture} 
\caption{Classification of the $u(2)$ subalgebras. Semisimple algebras are indicated in yellow. Elements, according to the definitions in the text, are indicated below each frame and the order (dimension) is given as a small number in the upper-right corner. }
\label{u2}
\ec
\end{figure}

Once a physical problem has been mapped into a Lie algebra, an interesting question is to study all the subalgebra chains that originate from that algebra, that might be called dynamical symmetries. More precisely a dynamical symmetry algebra occurs whenever a certain hamiltonian has been written as a linear combination of Casimir operators of all of the subalgebras in a chain \cite{IaAr,IL,Iac,FvI}. A neat discussion of the two chains of subalgebras for this case is given in Ref. \cite{FvI}, namely
$$
\begin{array}{lcl}
\begin{array}{cccc}
 I) & u(2) & \supset & u(1) \\
 & \mid &~& \mid \\
 &[N] &~& n_t
\end{array}
&~~~&
\begin{array}{cccccc}
 II) & u(2) & \supset & su(2)& \supset & so(2)  \\
 & \mid &~& \mid &~& \mid \\
 &[N] &~& j &~& \mu
\end{array} 
\end{array}
$$
Each one of them provides a basis that can be used to diagonalize the problem.
These two chains have a fundamental difference: one goes through the radical, the other through the semisimple Levi subalgebra
\footnote{While functionally independent invariants of semi-simple Lie algebra are well-known, the problem for non-semisimple Lie algebras is in general (but not in the present trivial case) much more complicated. See R.Campoamor-Stursberg, J.Phys. A {\bf 36} (2003).}.
Here the eigenvalues of two invariants are needed to label the basis states and while they are fixed in the first chain $\mid [N]n_t \rangle $, there is more freedom in the second as they can be chosen either as $\mid [N] j \rangle $ or $\mid j\mu \rangle$. The second form exhibits in a natural way the splitting of eigenstates with different third component of angular momentum and it has prevailed in the specialized literature.
The two subalgebras $u(1)$ and $so(2)$ are isomorphic, but when one is dealing with a complete classification of all subalgebras, some care must be taken. Although it is clear that the linear map that changes the sign of $g_4$ and leaves the rest unchanged, is actually swapping $u(1)$ and $so(2)$, {\it it does not send the whole reduction scheme of Fig. (\ref{u2}) into itself}. It can be easily proven that there is no Lie algebra homomorphism $\mathfrak{g} \rightarrow \mathfrak{g}$ mapping $u(1)$ into $so(2)$.

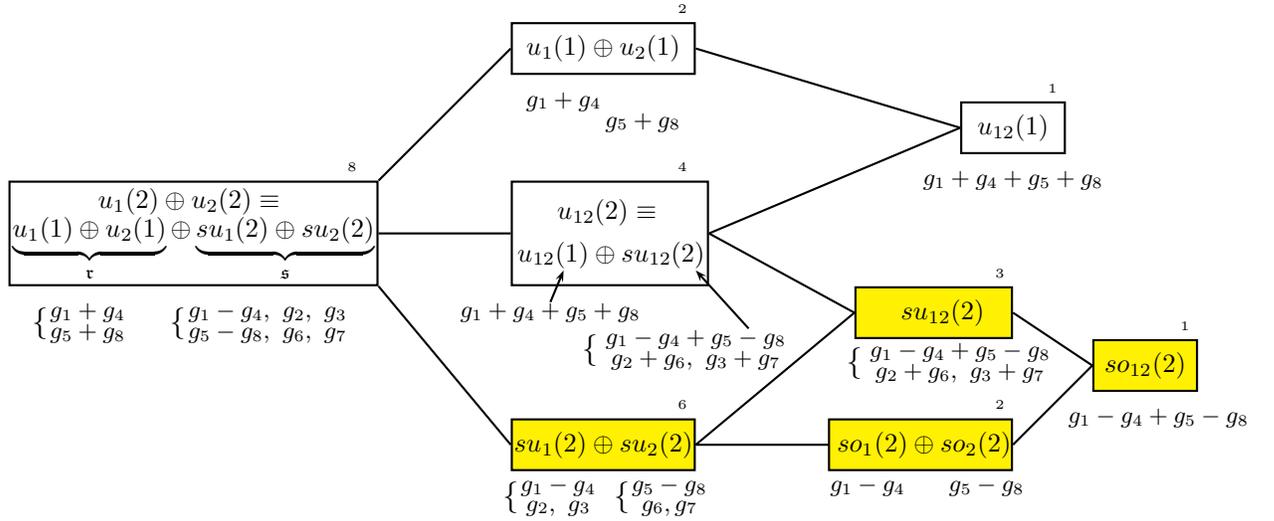
\begin{figure}[!t]
\bc
\begin{picture}(450,180)(0,0)
\psset{unit=1pt}
\psframe(0,120)(140,80)  \rput(130,125){\tiny 8}
\rput(70,112){$u_1(2)\oplus u_2(2)\equiv$ } 
\rput(70,94){ $\underbrace{u_1(1)\oplus u_2(1)}_{\mathfrak{r}}\oplus \underbrace{su_1(2)\oplus su_2(2)}_{\mathfrak{s}}$ }
\psline(140,120)(190,170)
\psline(140,100)(190,100)
\psline(140,80)(190,20)
\psframe(190,180)(260,160)  \rput(255,185){\tiny 2}
\rput(225,170){ $u_{1}(1)\oplus u_{2}(1)$ }
\psframe(190,120)(265,80)   \rput(255,125){\tiny 4}
\rput(228,108){$u_{12}(2)\equiv$ }
\rput(228,92){ $u_{12}(1)\oplus su_{12}(2)$ }
\psframe[fillcolor=yellow, fillstyle=solid](190,30)(260,10) \rput(255,35){\tiny 6}
\rput(225,20){ $su_{1}(2)\oplus su_{2}(2)$ }
\psline(260,20)(320,20)
\psline(260,20)(320,70)
\psline(265,100)(320,70)
\psline(265,100)(360,140)
\psline(260,170)(360,140)
\psframe[fillcolor=yellow, fillstyle=solid](320,80)(380,60) \rput(375,85){\tiny 3}
\rput(355,70){$su_{12}(2)$ }
\psframe[fillcolor=yellow, fillstyle=solid](310,30)(380,10)  \rput(375,35){\tiny 2}
\rput(348,20){$so_{1}(2) \oplus so_{2}(2)$ }
\psline(380,20)(410,50)
\psline(380,70)(410,50)
\psframe[fillcolor=yellow, fillstyle=solid](410,60)(450,40) \rput(445,65){\tiny 1}
\rput(430,50){ $so_{12}(2)$ }
\psframe(360,150)(400,130)   \rput(395,155){\tiny 1}
\rput(380,140){ $u_{12}(1)$ }
\rput(12,66){\large $\{$}
\rput(30,70){\small $g_1+g_4$}\rput(30,62){\small $g_5+g_8$}
\rput(64,66){\large $\{$}
\rput(98,70){\small $g_1-g_4,~g_2,~g_3$}\rput(98,62){\small $g_5-g_8,~g_6,~g_7$}
\rput(205,70){\small $g_1+g_4+g_5+g_8$}\psline{->}(205,74)(210,86)
\rput(220,56){\large $\{$}\psline{->}(280,64)(260,86)
\rput(260,60){\small $g_1-g_4+g_5-g_8$}\rput(260,52){\small $g_2+g_6,~g_3+g_7$}
\rput(210,150){\small $g_1+g_4$}\rput(240,142){\small $g_5+g_8$}
\rput(380,120){\small $g_1+g_4+g_5+g_8$}
\rput(190,0){\large $\{$}
\rput(208,4){\small $g_1-g_4$}
\rput(208,-4){\small $g_2,~g_3$}
\rput(232,0){\large $\{$}
\rput(250,4){\small $g_5-g_8$}
\rput(250,-4){\small $g_6,g_7$}
\rput(325,4){\small $g_1-g_4$}
\rput(370,4){\small $g_5-g_8$}
\rput(320,51){\large $\{$}
\rput(360,55){\small $g_1-g_4+g_5-g_8$}\rput(360,47){\small $g_2+g_6,~g_3+g_7$}
\rput(435,30){\small $g_1-g_4+g_5-g_8$}
\end{picture} 
\caption{Classification of the $u_1(2)\oplus u_2(2)$ subalgebras. Semisimple algebras are indicated in yellow. Elements, according to the definitions in the text, are indicated below each frame and the order (number of elements) is given as a small number in the upper-right corner. A total of 5 chains can be identified.}
\label{u2u2}
\ec
\end{figure}

\section{$u(2)\oplus u(2)$}
\label{sec-u2u2}
The case of the coupling of two independent systems of $s$ and $t$ bosons has also been studied in detail and applied to the case of vibrations of X-Y-X type molecules (such as water or carbon dioxide) \cite{FvI}. The first and second sets of bosons, indexed as 1 and 2, are associated with the two bonds of the molecule and the type, $s$ or $t$ are used to describe normal modes of motion. Although these kinds of models are usually very schematic, they catch the essential physics and they express it into an elegant mathematical formalism that provides simple symmetry-inspired energy formulas.
We repeat here the previous analysis to classify the possible subalgebra chains that might occur. As said there are two copies of the operators (\ref{op}) that close into the direct sum algebra $u_1(2)\oplus u_2(2)$. Let's divide them in the following way: $u_1(2)=\{g_1,g_2,g_3,g_4\}$ and $u_2(2)=\{g_5,g_6,g_7,g_8\}$. They are both non-semisimple as well as their direct sum, therefore one might apply the Levi-Malcev decomposition to get a radical $\mathfrak{r}=\{g_1+g_4, g_5+g_8\}$ and a Levi subalgebra $\mathfrak{s}=\{g_1-g_4,g_2,g_3, g_5-g_8,g_6,g_7\}$.
All the different subalgebras are given in Fig. \ref{u2u2} and they can be divided into their nilpotent or semisimple components, whenever it is due. 
By a careful examination of the figure, one can see that the algebra $u_{12}(1)$ cannot be included into $su_{12}(2)$ by any possible rearrangement of 
the elements. In fact the operator $\hat n_t= t_1^\dag t_1 +t_2^\dag t_2$ cannot be found as a subalgebra of ${\hat J}_{12,\rho}$ (with $\rho=x,y,z$) \footnote{At the light of this statement the chain (2.29) of Ref. \cite{FvI} must be reinterpreted: as the authors notice quite correctly, there is an isomorphism between $u(1)$ and $so(2)$, that they exploit in order to calculate the reduction rules, but the chain (2.29) is in fact just a copy of (2.30) with a different choice of the $so(2)$ generator, rather than a chain on its own (one can make a circular
permutation of the elements of the yellow algebras to get this). This is irrelevant as long as one concentrates 
on the physical applications. It amounts to a change of the label name that, for what has been said, cannot nevertheless count the number of $t$ bosons.}

\section{$u(3)$}
\label{sec-u3}
\begin{figure*}[!t]
\bc
\begin{picture}(400,160)(0,-5)
\psset{unit=1pt}
\psframe(0,100)(70,60)  \rput(65,105){\tiny 9}
\rput(35,90){$u(3)\equiv$ } 
\rput(35,72){ $\underbrace{u(1)}_{\mathfrak{r}}\oplus \underbrace{su(3)}_{\mathfrak{s}}$ }
\rput(20,55){\small $g_1$}
\rput(50,55){\small $g_2 \cdots g_9$}
\psline(70,100)(120,150)
\psline(70,90)(120,110)
\psline(70,60)(120,30)

\psframe(120,160)(210,140)  \rput(205,165){\tiny 4}
\rput(165,152){ $u(2)\equiv u(1)\oplus su(2)$ } 
\rput(202,144){\tiny [1,1]} 
\rput(160,135){\small $g_1$}
\rput(195,135){\small $g_5,g_3,g_9$}
\psframe(120,120)(210,100)   \rput(205,125){\tiny 4}
\rput(165,112){$u(2)\equiv u(1)\oplus su(2)$ }
\rput(202,104){\tiny [2,2]} 
\rput(160,95){\small $g_1$}
\rput(195,95){\small $g_2,g_3,g_4$}
\psframe[fillcolor=yellow, fillstyle=solid](120,40)(210,20) \rput(205,45){\tiny 8}
\rput(165,30){ $su(3)$ }
\rput(165,15){\small $g_2 \cdots g_9$}

\psline(210,30)(280,10)
\psline(210,30)(280,50)
\psline(210,110)(280,10)
\psline(210,150)(280,50)
\psline(210,110)(280,130)
\psline(210,150)(280,130)
\psframe(280,140)(320,120) \rput(315,145){\tiny 1}
\rput(300,130){$u(1)$ }
\rput(300,115){\small $g_1$}
\psframe[fillcolor=yellow, fillstyle=solid](280,60)(325,40)  \rput(320,65){\tiny 3}
\rput(300,50){$su(2)$ } 
\rput(317,45){\tiny [1,1]} 
\rput(300,35){\small $g_5,g_3,g_9$}
\psframe[fillcolor=yellow, fillstyle=solid](280,20)(325,0)  \rput(320,25){\tiny 3}
\rput(300,10){$su(2)$ }
\rput(317,5){\tiny [2,2]} 
\rput(300,-5){\small $g_2,g_3,g_4$}

\psline(325,10)(360,30)
\psline(325,50)(360,30)
\psframe[fillcolor=yellow, fillstyle=solid](360,40)(400,20) \rput(395,45){\tiny 1}
\rput(380,30){ $so(2)$ }
\rput(380,15){\small $g_3$}
\end{picture} 
\caption{Classification of the $u(3)$ subalgebras. Semisimple algebras are indicated in yellow. Elements, according to the definitions in the text, are indicated below each frame and the order (dimension) is given as a small number above the upper-right corner. Weighted Dynkyn diagrams are indicated in the lower-right corner: in the case of non-semisimple algebras they mark the semisimple part. A total of 6 chains can be identified.}
\label{u3}
\ec
\end{figure*}
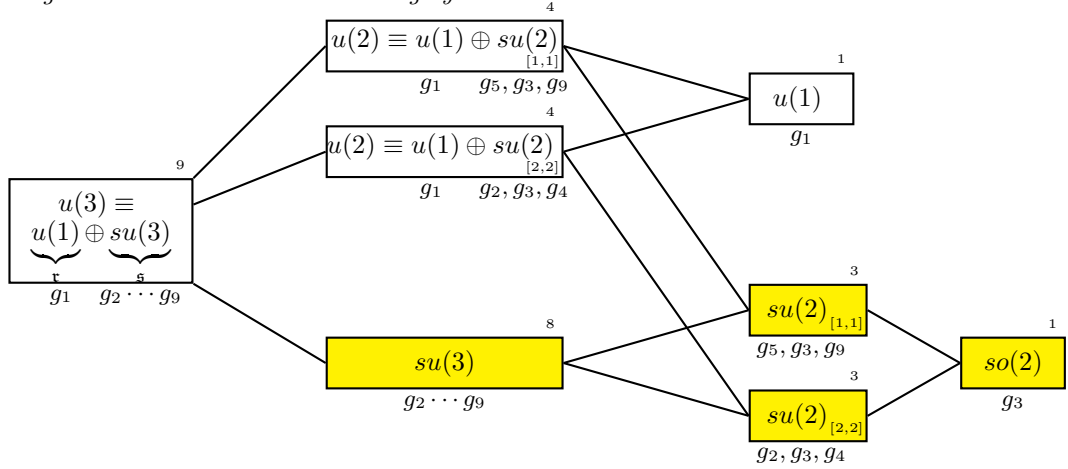

The next and most interesting step is to investigate the algebra $u(3)$, that is built upon bilinear combinations of three operators. These can be chosen either i) as three scalar bosons, that might be physically interpreted as appropriate combinations of the three Cartesian coordinates and momenta in a three dimensional harmonic oscillator, or ii) as the three components of a vector boson or finally iii) as a scalar boson plus two, so-called, circular bosons, that are infact the component of a spinor \cite{Iac}. Each construction has practical applications, especially to the study of molecular spectra.
We study here the construction ii) made up in terms of a $p$ $(\ell=1)$ boson, because it is necessary for the construction of the Vibron model \cite{IL,FvI,Oss}.
The operators $p^\dag_\mu$ and $\tilde p_\mu$ with $\mu=-1,0,1$ transform as the $\ell=1$ representation of the rotation group. They satisfy the usual boson commutation relations:
\be
[p_\mu,p_\nu^\dag]=\delta_{\mu\nu} \qquad[p_\mu,p_\nu]=[p_\mu^\dag,p_\nu^\dag]=0 \;.
\ee
The following nine bilinear operators (\ref{ts}) built from tensor couplings (Racah form) close into the $u(3)$ Lie algebra.
\be
\begin{array}{ccccc} 
\fl &&g_1=[p^\dag \times \tilde p]^{(0)}_{0}  && \\
\fl &g_2=[p^\dag \times \tilde p]^{(1)}_{-1} & g_3=[p^\dag \times \tilde p]^{(1)}_{0} & g_4=[p^\dag \times \tilde p]^{(1)}_{1} & \\
\fl g_5=[p^\dag \times \tilde p]^{(2)}_{-2} & g_6=[p^\dag \times \tilde p]^{(2)}_{-1} & g_7=[p^\dag \times \tilde p]^{(2)}_{0} & g_8=[p^\dag \times \tilde p]^{(2)}_{1} & g_9=[p^\dag \times \tilde p]^{(2)}_{2} 
\end{array} 
\label{ts}
\ee
where one can identify a scalar $\hat N=\sqrt{3}g_1$, a vector $\hat L_\kappa=\sqrt{2}[p^\dag \times \tilde p]^{(1)}_{\kappa}$ (3 components) and a quadrupole tensor $\hat Q_\kappa=[p^\dag \times \tilde p]^{(2)}_{\kappa}$  (rank 2, 5 components) to be further discussed below \cite{Iac}.
One can apply the same arguments as in the preceding section, namely the Levi-Malcev decomposition and the classification of $A_1$ subalgebras in terms of weighted Dynkin diagrams, to show that the lattice of subalgebras of $u(3)$ takes the form displayed in Fig. (\ref{u3}). The Levi-Malcev decomposition is clear from the division into white and yellow blocks  in Fig. (\ref{u3}) and the scalar operator $g_1$ is actually always responsible for the radical part of the classification. From Figs. (\ref{u2}) and (\ref{u3}) one can see that the whole scheme can be divided in two parallel sheets, a {\it semisimple sheet} (our terminology), containing all the semisimple Lie algebras, and a {\it non-semisimple sheet}, containing an exact copy of the structure of the lower one, where each algebra is multiplied by the radical. Of course $u(n)$ corresponds to $su(n)$ and so on, each corresponding pair is also connected by an inclusion relation from top to bottom with the exception of one dimensional subalgebras.
The semisimple part starts at $su(3)$, that is $A_2$ in Cartan's notation
\footnote{Note that the present yellow part is equivalent to the right scheme (7.80) of Ref. \cite{Iac}. In the left scheme the notation of unitary algebras should be properly replaced by special unitary.}: this algebra has two types of $A_1$ subalgebras labeled by different weighted Dynkin diagrams, $[1,1]$ and $[2,2]$ respectively. 
Each of them is an entire conjugacy class of triplets of operators with the same WDD of which we choose just one representative (usually the simplest available or the one that has already been incorporated into an established model).
The second one is the usual algebra of angular momentum, whose components are the components of the rank 1 tensor in Eq. (\ref{ts}). Indeed the three operators $\hat L_\kappa=\sqrt{2}[p^\dag \times \tilde p]^{(1)}_{\kappa}$ satisfy the angular momentum algebra. This algebra and all the chains passing through it have been described by Iachello and the physical reasoning underlying this rightful choice is that the quantum mechanical description of molecular systems and the basis states associated to the chain demand conservation of the angular momentum.

It turns out, however, and this fact was unknown or mostly unnoticed till now, that the $A_1$ algebra labeled by $[1,1]$ is made up of three objects that are not components of a vector, {but that, nevertheless, have commutation relations that formally identify them as an angular momentum algebra}
\be
[\hat W_+, \hat W_-]=2W_0 \qquad [\hat W_0, \hat W_{\pm}]=\pm \hat W_{\pm}
\ee
where the operators are defined as $\{ \hat W_-=g_5, W_0=g_3/\sqrt{2}, W_+=g_9 \}$. In usual terminology they are the $\hat Q_{\pm 2}$ components of the quadrupole tensor and the $\hat L_0$ component of the angular momentum. Historically only other two theoretical works have introduced something of this sort: Chen and Arima \cite{CA} discuss the origin of cylindrical bosons within the Interacting Boson Model, where they introduce the $\Delta$ spin that is built upon the highest and lowest components of the quadrupole tensor plus the zero component 
of the angular momentum, the difference being that their operators are made up of $s$ and $d$ bosons, while ours are made up of $p$ bosons 
\footnote{Notice that the notation in the mentioned paper may be somewhat confusing: they invoke the group chain $SU(3) \supset SU(2) \otimes U(1)$, while they are actually using either the $u(3) \supset u(2)[1,1] \otimes u(1)$ or rather $su(3) \supset su(2)[1,1]$. This kind of ambiguities are fully resolved in our classification scheme.}; Elliott  discusses, in the fundamental Ref. \cite{Ell}, a similar algebraic structure in the context of the collective motion in the nuclear shell model.

Now the crucial point is that once we have an angular momentum algebra, say $\hat J$, (to be replaced either by $\hat L$ or by $\hat W$) we can define spherical tensors with respect to {\em that} algebra by means of Eq. (\ref{sto}).
We will use the (somewhat tedious, but necessary) name of "J-tensors" to refer to tensors with respect to a particular "J-set".
The whole $u(3)$ algebra is made up by an L-scalar, an L-vector and and L-tensor of rank 2, as outlined above after Eq. (\ref{ts}), with respect to the L-set. In this L-set, both $p^\dag$ and $\tilde p$ transform as L-vectors. 
The analysis of the elements of $u(3)$ with respect to the W-set reveals instead that, together with the obvious W-vector given by $\hat W$ itself, we have two W-scalars and two W-spinors. This might come as a surprise since, in general, we are not expecting spin $1/2$ operators to arise in purely bosonic models (although it is clear that these algebras might have spinor representations). The proof goes as follows:\\
a) $g_1$ and $g_7$ commute with all components of $\hat W$: they are therefore scalars.\\
b) the objects $sp_1=\{g_6,g_4\}$ and $sp_2=\{-g_8,-g_2\}$ (where the components have the $\{-,+\}$ order) have the following commutation relations with $\hat W_0$:
\be
[\hat W_0, sp_i^\pm] = \pm \frac{1}{2} sp_i^\pm 
\ee
with $i=1,2$ and therefore they are the $-1/2$ and $+1/2$ components of a spinor.\\
c) the Lie products with $\hat W_+$ (resp.  $\hat W_-$) terminate after two steps:
\bea 
[\hat W_+, sp_i^-] =  sp_i^+ & \quad [\hat W_+, sp_i^+] =  0 \\ \nonumber
[\hat W_-, sp_i^+] =  sp_i^- & \quad [\hat W_-, sp_i^-] =  0 \nonumber
\eea
with $i=1,2$. The coefficients are compatible only with $k=1/2$ in Eq. (\ref{sto}), therefore we have found two spin$-1/2$ spinors. We will reserve the name of Chen-Arima spinors for spinors arising in bosonic models, because these authors have provided the first case.
Notice that one can always form other scalars from tensor coupling of Chen-Arima spinors obtaining in our case:
\be
[sp_1 \times sp_1]^{(W=0)}_0=-[sp_2 \times sp_2]^{(W=0)}_0= -\frac{\sqrt{3}}{2} g_7 
\ee
that is a W-scalar consistent with the observation a) made above.

Another interesting question is the fate of the vectors $p^\dag$ and $\tilde p$. They don't satisfy good tensorial properties with respect to the W-set.
Their role is replaced by other two objects $V=\{\sqrt{2}p^\dag_{-1}p^\dag_{-1},2p^\dag_{-1}p^\dag_{1},\sqrt{2}p^\dag_{1}p^\dag_{1} \}$ and 
$U=\{\sqrt{2}\tilde p_{-1} \tilde p_{-1},2\tilde p_{-1}\tilde p_{1},\sqrt{2}\tilde p_{1}\tilde p_{1} \}$ (with components ordered as $-1,0,1$) that 
are found to be good W-vectors by following the same argument outlined above. Their expressions, that consist in creating or annihilating two bosons 
at the same time are in tune with the definitions of the components of the W-spin.

\section{$u(4)$}
\label{sec-u4}
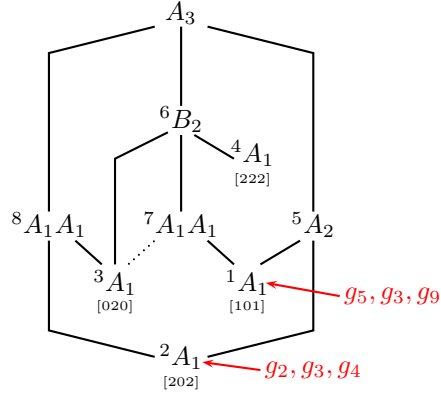
\begin{figure}[!t]
\bc
\begin{picture}(100,140)(0,0)
\psset{unit=1pt}
\rput(50,140){$A_3$}
\rput(50,100){$^6B_2$}
\rput(100,60){$^5A_2$}
\rput(0,60){$^8A_1A_1$}
\rput(50,60){$^7A_1A_1$}
\psline{-}(40,135)(0,125)(0,65)
\psline{-}(60,135)(100,125)(100,65)
\psline{-}(50,135)(50,105)
\psline{-}(50,94)(50,65)
\psline{-}(55,94)(70,85)
\rput(25,38){$^3A_1$} \rput(25,29){\tiny $[020]$}
\rput(75,38){$^1A_1$} \rput(75,29){\tiny $[101]$}
\rput(77,87){$^4A_1$}  \rput(77,77){\tiny $[222]$}
\rput(50,10){$^2A_1$} \rput(50,0){\tiny $[202]$}
\psline{-}(10,54)(20,45)
\psline[linestyle=dotted,dotsep=2]{-}(40,54)(30,45)
\psline{-}(45,95)(25,85)(25,45)
\psline{-}(60,54)(70,45)
\psline{-}(95,54)(80,45)
\psline{-}(0,54)(0,20)(40,10)
\psline{-}(100,54)(100,20)(60,10)
\rput(100,5){\red $g_2,g_3,g_4$} \psline[linecolor=red]{->}(80,5)(58,8)
\rput(130,33){\red $g_5,g_3,g_9$} \psline[linecolor=red]{->}(110,33)(82,38)
\end{picture} 
\caption{Classification of Lie subalgebras of $A_3$ adapted to the present case (arbitrary upper left indices). Three-dimensional subalgebras of type $A_1$ ($sl_2-$triples) represent conjugacy classes, labeled by the corresponding WDD in square brackets. In theory all inclusions are possible, but in practice, in our particular realization, the dotted one is not possible simultaneously with all the others, although the direct inclusion  $^6B_2 \supset ^3A_1$ is valid. Elements of selected subalgebras are indicated in red.}
\label{su4}
\ec
\end{figure}

This is the all-important case of the Vibron model \cite{IL,Iac,FvI,Oss}, that is built upon $s$ and $p$ bosons. Together with the nine generators in Eq. (\ref{ts}) one needs other seven generators containing $s$ to close the algebra, namely:
\be \begin{array}{ccccc} 
g_{10}=[s^\dag \times \tilde p]^{(1)}_{-1} && g_{11}=[s^\dag \times \tilde p]^{(1)}_{0} && g_{12}=[s^\dag \times \tilde p]^{(1)}_{1} \\
g_{13}=[p^\dag \times \tilde s]^{(1)}_{-1} && g_{14}=[p^\dag \times \tilde s]^{(1)}_{0} && g_{15}=[p^\dag \times \tilde s]^{(1)}_{1} \\
&&g_{16}=[s^\dag \times \tilde s]^{(0)}_{0} && \\
\end{array}
\label{ts2}
\ee
The radical of the $u(4)$ algebra is given by $\hat N=\sqrt{3}g_1+g_{16}$ that is the total number of bosons operator. The semisimple sheet consists (see Fig.\ref{su4}) of the seven chains that originate from $A_3$ end end up in one of the four possible $A_1$, the conjugacy classes of which are labeled by the WDD: $[101], [202], [020]$ and $[222]$. In the Vibron model only two such chains, the ones passing through the standard angular momentum subalgebra $[202]$ have been studied in great detail so far. 

The well-known chains of the Vibron model are the most external paths of Fig. \ref{su4} connecting $A_3$ and $A_1[202]$ or, for the sake of clarity (in either usual and Cartan's labels):
\be
 su(4) \supset \left\{
\begin{array}{l}  
~su(3)\\ \\
~so(4)  
\end{array} 
\right\}
\supset  so(3) 
\qquad
A_3 \supset \left\{
\begin{array}{l}  
~^5A_2\\ \\
~^8A_1A_1  
\end{array} 
\right\}
\supset  ~^2A_1
\ee
that correspond to the nonrigid and rigid rotovibrator limits of the Vibron model.
In the complete classification four classes of $A_1$ subalgebras are present and one has therefore four different "angular momenta" that can be used as J-sets to define spherical tensors. The W-angular momentum described in the previous section for $u(3)$ is also present here and it forms the $^1A_1$ algebra with WDD $[101]$. With respect to this W-set, the whole algebra $su(4)$ amounts to the W-vector $\hat W$, to four W-scalars $\hat s_1, \cdots, \hat s_4=$ $g_1,g_7,g_{11},g_{14}$ and to  four Chen-Arima W-spinors $\hat {sp}_1, \cdots, \hat {sp}_4$ $=\{g_6,g_4\},\{-g_8,-g_2\},\{g_{10},g_{12}\},\{-g_{15},-g_{13}\}$. The scalars built with tensor couplings of $\hat sp_3$ with itself and $\hat sp_4$ with itself are identically zero, therefore one can define two new W-spinors $\hat sp_3'=$ $\{-g_2-g_{10}-g_{13}, g_8-g_{12}+g_{15}\}$ and $\hat sp_4'=$ $\{g_6+g_{10}-g_{13}, g_4+g_{12}+g_{15}\}$ as linear combinations.  Their couplings give good W-scalars and they are also needed as elements of $B_2$.

\begin{table}[!t]
\bc
\begin{tabular}{cccc}\hline 
operator & def$^{rank}$ & alt.def.& components \\ \hline
$\hat n'$ &  $ [s^\dag \times \tilde s]^{0}_0- \sqrt{\frac{1}{3}}[p^\dag \times \tilde p]^{0}_0 $ & $g_1$ & 1  \\
$\hat L_\mu /\sqrt{2}$  & $ [p^\dag \times \tilde p]^{1}_\mu $ & $g_2,g_3,g_4$ & 3 \\
$\hat Q_\mu$  & $ [p^\dag \times \tilde p]^{2}_\mu $   & $g_5,\cdots, g_9 $ & 5 \\
$\hat D_\mu$  & $ i[p^\dag \times \tilde s+ s^\dag \times \tilde p]^{1}_\mu $ & $g_{10},g_{11},g_{12} $& 3 \\
$\hat D_\mu'$ & $ [p^\dag \times \tilde s- s^\dag \times \tilde p]^{1}_\mu $  & $g_{13},g_{14},g_{15} $& 3 \\ \hline
\end{tabular} \ec
\caption{Definitions and tensorial character of the operators forming the semisimple $A_3$ algebra with respect to the (standard) set forming the $^2A_1$ algebra with $[202]$ WDD. The third column features alternative definitions for algebra elements that correspond to the second column with the following ordering $\mu=-k,..,0,..,k$, where $k$ is the rank.}
\label{tab1}
\end{table}

\begin{table}[!t]
\bc
\begin{tabular}{ccc}\hline 
operator & definition & components \\ \hline
$\hat s_1, \cdots, \hat s_4$ &  $ g_1,g_7,g_{11},g_{14} $ & 1 (each) \\
$\hat W_\mu$  & $ -g_5/\sqrt{2}, g_3/\sqrt{2}, g_9/\sqrt{2} $ & 3 \\
$\hat {sp}_1, \cdots, \hat {sp}_4 $ & $ \{g_2,g_8\},\{g_6,g_4\},\{g_{10},g_{12}\},\{g_{13},g_{15}\} $ & 2 (each)\\ \hline
\end{tabular}
\ec
\caption{Definitions and tensorial character of the operators forming the $A_3$ algebra with respect to the non-standard set forming the $^1A_1$ algebra with $[101]$ WDD (that can be obtained from the vector in the first line). }
\label{tab2}
\end{table}

\begin{table}[!t]
\bc
\begin{tabular}{ccc}\hline 
operator & definition & components \\ \hline
$\hat Y_\mu$& $  i(g_2 + g_6)/\sqrt{2} +g_{10},(g_1 + \sqrt{2}g_3 + \sqrt{2/3}g_7 + 2ig_{11})/2, i\sqrt{2}g_4 + g_{12}+g_{15} $ & 3 \\
$\hat B_\mu$& $ 2\sqrt{2}g_9, \sqrt{2}i(g_4-g_8)-2g_{12}, 2i\sqrt{2}g_{11}  $ & 3 \\
$\hat M_\mu$& $ -g_1-\sqrt{2/3}g_7+i(g_{14}-g_{11}), ig_6+(g_{13}-g_{10})/\sqrt{2}, g_5 $ & 3 \\ 
$\hat R_\mu$& $ -i(g_{4}+g_{8})/\sqrt{2}+g_{12}, -g_3/2-\sqrt{3/8}g_7-ig_{11}, i(g_2+g_6)/\sqrt{2} $ & 3 \\ 
$\hat S_\mu$& $ -i(g_{4}/g_{8})/\sqrt{2}+g_{15},  -g_1/\sqrt{2}-g_3/2+\sqrt{1/12}g_7-ig_{11}, -g_{10}$ & 3 \\ \hline
\end{tabular}
\ec
\caption{Definitions and tensorial character of the operators forming the $A_3$ algebra with respect to the non-standard set forming the $^3A_1$ algebra with $[020]$ WDD (that can be obtained from the vector in the first line). }
\label{tab3}
\end{table}

\begin{table*}[!t]
\bc
\begin{tabular}{ccc}\hline 
oper. & definition & comp. \\ \hline
$\hat T^{(1)}_\mu$ &  $ (g_{2}-7g_4-7g_6-g_8+3g_{13})/2, -\frac{3}{2}g_1-\sqrt{2}g_3-2g_5, g_2+g_{10}+g_{12}$ & 3 \\
$\hat T^{(2)}_\mu$  & $ -6\sqrt{2}(2g_3+\sqrt{2}g_5-\sqrt{2}g_9+g_{14}), -3(g_2+g_4+g_6-g_8+g_{13}), $ & 5\\
 & $ \sqrt{3/2}g_1-\sqrt{3}g_3-\sqrt{6}g_5+g_7,(g_6-g_2)/2+g_{10}+g_{12},(g_5+\sqrt{2}g_{11})/4 $ &  \\
$\hat T^{(3)}_\mu$ & $ \frac{-72}{\sqrt{30}}(g_{13}-g_{15}),\frac{-12}{\sqrt{10}}(2g_2+\sqrt{2}g_5-\sqrt{2}g_9-g_{14}), \frac{-6}{5\sqrt{2}}(3g_2-g_4-g_6-3g_8-g_{13}),$ & 7 \\
& $ (-\sqrt{6}g_1+2(\sqrt{3}g_3+\sqrt{6}g_5+5g_7))/10, (g_2+5g_6-4g_{10}-4g_{12})/(10\sqrt{2}),$ &\\ 
&$(g_5/\sqrt{2}-g_{11})/\sqrt{40}, -g_{10}/\sqrt{120} $ & \\ \hline
\end{tabular}
\ec
\caption{Definitions and tensorial character of the operators forming the $A_3$ algebra with respect to the non-standard set forming the $^4A_1$ algebra with $[222]$ WDD (that can be obtained from the vector in the first line). }
\label{tab4}
\end{table*}

We have summarize in Tables \ref{tab1}, \ref{tab2}, \ref{tab3} and \ref{tab4} the tensor analysis of the whole algebra $A_3$ with respect to $^2A_1[202]$, $^1A_1[101]$, $^3A_1[020]$ and $^4A_1[222]$ respectively. So, for example, $A_3$ is made up of 5 vectors with respect to $^3A_1[020]$ (among which the defining vector $Y$ itself), while it is made up of a vector (the defining vector $T^{(1)}$), a quadrupole and an octupole tensor in the latter case.

\section{Applications in physics}
\label{appl}
One might think of writing AMNC hamiltonians with the dynamical symmetry based on the chain $A_3 \supset B_2 \supset {^7A_1A_1}\supset  {^1A_1}$ or 
\be
\left|
\begin{array}{ccccccc}
 su(4) \sim so(6)& \supset & so(5)& \supset &  so(4)    & \supset & so(3) \\
\mid            &     ~   & \mid &   ~      & \mid     &    ~     & \mid  \\
N            &     ~   &  t   &    ~     &   u      &    ~     &  w   \\
\end{array} 
\right\rangle
\ee
where the labels in the last row are connected with the eigenvalues of the quadratic Casimir operators, namely
\be
\begin{array}{ccc}
\langle C_2(so(6)) \rangle &=&N(N+4) \\
\langle C_2(so(5))\rangle &=&t(t+3) \\
\langle C_2(so(4))\rangle &=&u(u+2)\\
\langle C_2(so(3))\rangle &=&w(w+1) \;.
\end{array} 
\ee 
A hamiltonian with this dynamical symmetry is obtained by linear combination of the above Casimir operators as:
\be
H=\alpha C_2(so(6))+\beta C_2(so(5)) + \gamma C_2(so(4)) +\delta C_2(so(3)) \;.
\ee
The resulting energy formula for symmetric representations is 
\be
E=\alpha N(N+4)+\beta t(t+3) + \gamma u(u+2) +\delta w(w+1)
\label{ener}
\ee
with branching rules: $t=0,\cdots, N$, $u=0,\cdots, t$ and $w=0,\cdots, u$.
Although Eq. (\ref{ener}) does not look like a good choice for diatomic molecular spectra, one can anyway use the basis states $\mid N,t,u,w \rangle$ (actually a Gelfand-Tsetlin pattern for orthogonal algebras) to diagonalize hamiltonians based on the full spectrum generating algebra, with the {\it proviso} that, while $N$ is the total boson number as in the Vibron model, the labels $v,u$ and $w$ {\em do not conserve angular momentum, but rather conserve w}. Earlier on one would have doubted the value of using this basis for the Vibron model, due to the difficulty in giving a precise physical meaning to the labels. One advantage is the ease of writing the branching rules (and this might imply the absence of missing labels in higher order algebras).
Several other AMNC dynamical symmetries, one for every possible path in Fig. \ref{su4} not ending in the $[202]~~^2A_1$ subalgebra, can be invoked: they provide {\it at least} new diagonalization schemes and possibly applications to quantum many-body systems.

\section{Conclusions}
\label{conc}
We have shown that the use of i) Levi-Malcev decomposition and ii) theory of weighted Dynkin diagrams allow a thorough classification of algebraic models arising in bosonic models of quantum mechanics providing i) a way of separating out semisimple from non-semisimple subalgebras and ii) a neat classification of all possible conjugacy classes of three dimensional subalgebras ($A_1$) respectively. Well-known algebraic models, either used as conventional or pedagogical toy models or actually applied to real systems, usually adopt subalgebra chains that end up in the standard angular momentum algebra. While this is a perfectly reasonable choice, we have shown that, even within the sets of bilinear operators that are commonly defined on textbooks for a given algebra, one can "fill the gaps", i.e. write basis elements for the whole reduction scheme. In particular the elements of additional $A_1$, having different WDD, can be used to define new angular momenta operators, with respect to which one can redefine tensors and give to the whole algebra elements a different tensorial character.
With respect to one of these new angular momenta, it is found that other operators behave like spin-1/2 objects, a fact that was most surprising at first. This was hinted at by the old works of Elliott \cite{Ell} and Chen-Arima \cite{CA} and our paper provides a more complete collocation for their findings. In particular Chen and Arima have found spinors arising within the Interacting Boson Model of the nucleus that is completely bosonic (built upon $s$ and $d$ bosons). We argue that they have found the $A_1[11011]$ subalgebra of $su(6)$, the analog of our $A_1 [101]$ subalgebra of $su(4)$.
Another aspect worth mentioning again is that, although hamiltonian displaying dynamical symmetries based on AMNC chains might be unphysical, the basis states associated to them provide an alternative (maybe easier) basis for diagonalization of complex hamiltonians.

We believe that there might be other insightful discoveries or advantages awaiting in the still uninspected angular momentum non conserving chains of algebraic models.

\section{Acknowledgments}
We thank Luigi Scorzato (ECT*) for several interesting discussions.
L.F. thanks all the participants at the workshop held in Huelva (Spain) 17-18/05/2010 for the lively discussion that provided fruitful insight. L.F. acknowledges F.Iachello and A.Leviatan for several interesting comments borne out of the talk \cite{FdG} given at the ICGTMP'28 conference held in Newcastle-upon-Tyne (July 2010).

\end{document}